\documentclass[
 aps, 
 pra,
 twocolumn,
 reprint,
 noeprint,
 superscriptaddress,
 ]{revtex4-2}

\usepackage{amsmath}
\usepackage{amssymb}
\usepackage{graphicx}
\usepackage{bm}
\usepackage{color}
\usepackage{subfigure}
\usepackage{pdfpages}
\usepackage{pgffor}
\usepackage{ulem}
\usepackage{mathrsfs}
\usepackage[colorlinks, 
            linkcolor=blue, 
            anchorcolor=blue, 
            citecolor=blue,
            ]{hyperref}

\newcommand{\Rb}[1]{$^{#1}\rm{Rb}$}
\newcommand{\Na}[1]{$^{#1}\rm{Na}$}
\newcommand{\Li}[1]{$^{#1}\rm{Li}$}

\def\Hz{{{\rm Hz}}}	
\def\kHz{{{\rm kHz}}}	
\def\THz{{{\rm THz}}}		
\def\MHz{{{\rm MHz}}}	
\def\GHz{{{\rm GHz}}}

\makeatletter
\AtBeginDocument{\let\LS@rot\@undefined}
\makeatother

\begin{document}

\title{Controlling atomic spin-mixing via multiphoton transitions in a cavity}

\author{Ming Xue}
\email{mxue@nuaa.edu.cn}
\affiliation{College of Physics, Nanjing University of Aeronautics and Astronautics, Nanjing 211106, China}
\affiliation{Key Laboratory of Aerospace Information Materials and Physics (NUAA), MIIT, Nanjing 211106, China}

\author{Xiangliang Li}
\affiliation{Beijing Academy of Quantum Information Sciences, Xibeiwang East Road, Beijing 100193, China}

\author{Wenhao Ye}
\affiliation{State Key Laboratory of Low Dimensional Quantum Physics, Department of Physics, Tsinghua University, Beijing 100084, China}

\author{Jun-Jie Chen}
\affiliation{Equity Derivatives Business Line, CITIC Securities, Liangmaqiao Road, Beijing 100026, China}

\author{Zhi-Fang Xu}
\email{xuzf@sustech.edu.cn}
\affiliation{Shenzhen Institute for Quantum Science and Engineering and Department of Physics,
Southern University of Science and Technology, Shenzhen 518055, China}

 \author{Li You}
 \email{lyou@tsinghua.edu.cn}
 \affiliation{Beijing Academy of Quantum Information Sciences, Xibeiwang East Road, Beijing 100193, China}
 \affiliation{State Key Laboratory of Low Dimensional Quantum Physics, Department of Physics, Tsinghua University, Beijing 100084, China}


\begin{abstract} 
We propose to control spin-mixing dynamics in a gas of spinor atoms, via 
the combination of two off-resonant Raman transition pathways,  
enabled by a common cavity mode and a bichromatic pump laser.
The mixing rate, which is proportional to the synthesized spin-exchange interaction strength, 
and the effective atomic quadratic Zeeman shift (QZS), can both be tuned by changing the pump laser parameters.
Quench and driving dynamics of the atomic collective spin
are shown to be controllable on a faster time scale than 
in existing experiments based on inherent spin-exchange collision interactions.
The results we present open a promising avenue for exploring spin-mixing physics of atomic ensembles accessible in current experiments.
\end{abstract}
\maketitle

\section{Introduction}
Establishing quantum entanglement between two parties is crucial to 
quantum technology\,\cite{raimond01rmp,morris20prx}.
A direct approach for entanglement generation is based on coherent interaction between parties. 
To manipulate and protect quantum entanglement in a quantum many-body 
system, strong and precisely controllable quantum interaction is required.
Quantum phases with different entanglement properties can be realized by 
tuning relative strengths of competing interactions\,\cite{sachdev_2011}.

Between spinful atoms, spin-exchange interaction naturally arises 
when binary collision strengths differ for different total spin channels\,\cite{Ho1998,Law1998}. 
Coherent quantum spin-mixing dynamics, modeled by contact spin-exchange interaction 
between pairs of atoms, have been observed for spinor Bose-Einstein condensate (BEC) in all-optical trap experiments, 
using \Rb{87}, \Na{23}, \Li{7} atoms\,\cite{Stamper98,Stenger1998,Chang2004,Chang2005, Sadler2006,Choi20Li7}, 
as well as their mixtures\,\cite{DJW15PRL,Wang_2015,wang20heter}.
The resulting collective population oscillations among different spin states, 
enable the exploration of many-body physics related to spin degrees of freedom, 
e.g., spin squeezing\,\cite{Lucke2011,Hamley2012,Muessel2014scalable,hoang2016parametric}, 
generation of metrologically meaningful entangled states\,\cite{peise2015satisfying,Klempt2016PRL,Oberthaler2016PRL,Luo2017,Zou201715105}, 
and in probing nonequilibrium dynamics\,\cite{Saito2007Kibble,Oberthaler2015PRL, Anquez2015quantum, prufer2018observation,Xue2018,Schmied19pra}.
Any endeavor to establish tunable two-body
interaction is desirable for ground-state atoms in a condensate, 
where inherent atomic spin-exchange interaction is nominally weak\,\cite{Chang2004,Chang2005, Sadler2006,Choi20Li7}. 

Varying the density of particles could simply tune atomic interaction strength versus single-particle energy\,\cite{Greiner2002}.
More elaborate techniques like Feshbach resonance\,(FR) achieves the same at constant density\,\cite{Chin10RMP},
by tuning the scattering energy between two atoms through a nearby closed-channel 
molecular bound state\,\cite{Inouye1998,Stenger99,Blatt11optical,Wu12Optical}, 
which in some limiting cases can be viewed as inducing atom-atom interactions by coupling off-resonantly to their bound molecular state.
Such a picture extrapolates smoothly to the scenario of indirect interaction mediated by a
quantum channel, or more generally any intermediate bosonic quantum object.
The physical constituent of the channel can be an electromagnetic field mode in a cavity or photonic crystal\,\cite{Kapale05Cavity,Mottl2012, Aron16Photon,Hung2016,Zhang2017c,Masson2017,davis19photon,mivehvar19prl}, 
vibrational phonons in trapped ions\,\cite{Cirac04PRL,kim2010quantum,Britton2012} 
or opto(spin)-mechanical hybrid system\,\cite{16prlOptomechanical,20prlspinmechanical},
and atoms with dipole-dipole interactions limited to
excited Rydberg state manifold\,\cite{Bouchoule02spin,Henkel10PRLthree, Glaetzle15designing,Schauss1455,Zeiher2017PRX,fyang20prl,borish20transverse}.

Here in this work, we present a simple but efficient scheme for controlling spin-mixing dynamics
in spinor atomic gases using only optical fields.
Extending earlier studies\,\cite{Masson2017,davis19photon,Monika21Programmable}, we show that by using two $\sigma$-polarized laser fields 
in an atom-cavity system, the effective spin-exchange interaction between 
ground-state atoms and the effective atomic quadratic Zeeman shift (QZS) 
become tunable without requiring more complicated setups.
Besides synthesized spin-exchange interaction in spin-1 atoms as 
previously studied\,\cite{Masson2017,davis19photon,Monika21Programmable},
the ability to tune QZS in the present scheme provides a critical ingredient for realizing 
a rich variety of quantum phases\,\cite{Murata2007, uchino10pra} 
and for related quantum metrological applications of spinor atoms\,\cite{peise2015satisfying,Klempt2016PRL,Oberthaler2016PRL,Luo2017,Pezze19herald}.
The effective QZS can be easily tuned to compete with photon-mediated effective interaction 
by simply varying the differential laser detuning,
resulting in the formation of different quantum phases 
as well as for providing faster controlled dynamics\,\cite{Murata2007, uchino10pra, Luo2017}.
Thus our work goes beyond that of the analogously synthesized interactions in spin-1 atomic systems of earlier studies\,\cite{Masson2017,davis19photon,Monika21Programmable}, 
where only linear Zeeman shifts in external magnetic fields were considered
and QZSs were not tunable without additional dressing laser or microwave fields\,\cite{Masson2017,Monika21Programmable}.
By facilitating easy tuning of both the effective spin-exchange strength and 
QZS, our approach can be adapted to systems inside a significant bias magnetic field, 
while maintaining the desired interaction and the consequent spin-mixing dynamics.
\begin{figure}
  \centering
\includegraphics[width=1\columnwidth]{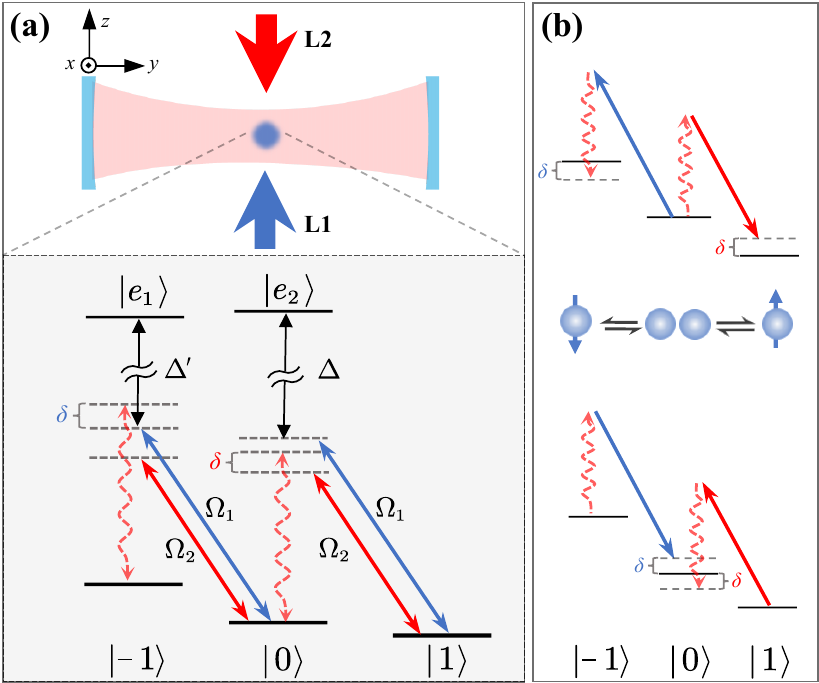}
\caption{(a) Atoms inside an optical cavity are pumped by two $\sigma_-$-polarized 
lasers with frequencies $\omega_1, \omega_2$ and Rabi frequencies $\Omega_1, \Omega_2$ from side, 
denoted as L1 (blue) and L2 (red solid lines), respectively.
Spin-1 atomic ground states ($|0\rangle, |-1\rangle$) are coupled to two excited states $|e_1\rangle,\,|e_2\rangle$ by 
a single cavity mode (red-wavy lines) with frequency $\omega_c$ and coupling strength $g_{i=1,2}$. 
{Cavity axis is along the $y$-direction with polarization axis along $z$, the direction of external magnetic field $B_z$.}
(b) Taking $\omega_1-\omega_2=2q$, the two four-photon Raman transition pathways illustrated
give rise to resonant atomic spin-exchange.}\label{fig:scheme} 
\end{figure}

\section{Model and Hamiltonian}
Our scheme is illustrated intuitively as in Fig.\,\ref{fig:scheme} for 
a cloud of spin-1 atomic gas or BEC tightly trapped inside an optical cavity.
To simplify our discussion, atom-light coupling is assumed spatially uniform,
which could be achieved by selective loading of atoms into a spatial lattice or alternatively by using a ring cavity\,\cite{Elsasser2004,Esslinger2007,Monika21Programmable}.
Atoms are pumped by two external $\sigma_-$-polarized lasers (labeled as L1 and L2 with respective 
frequencies $\omega_1$ and $\omega_2$) from the side and also are coupled to a single cavity mode (with frequency $\omega_c$).
The atomic level diagram contains two excited states $|e_1\rangle$ and $|e_2\rangle$, e.g., the $5P_{1/2}$ or $5P_{3/2}$ states for \Rb{87} atoms\,\cite{Esslinger2007}. 
In the lower pannel of Fig.\,\ref{fig:scheme}\,(a), L1 and L2 induce $\sigma$-transitions
$|0\rangle\leftrightarrow|e_1\rangle$ and $|1\rangle\leftrightarrow|e_2\rangle$
with coupling strengths $\tilde{\Omega}_{i=1,2}$ and $\Omega_{i=1,2}$ 
($\tilde{\Omega}_i=\Omega_i$ is assumed),
and the cavity field (wavy line) couples $\pi$-transitions $|0\rangle\leftrightarrow|e_2\rangle$ and $|\!-\!1\rangle\leftrightarrow|e_1\rangle$ 
with strengths $g_1$ and $g_2$, respectively.

Spinor BEC of $F=1$ ground state atoms, e.g., \Na{23} or \Rb{87} atoms with anti-ferromagnetic or ferromagnetic spin-exchange 
interactions have been studied extensively\,\cite{Sadler2006,Zhao2015a,Bookjans2011,Hoang2013,Luo2017,dalibard21prl}. 
Under single-mode approximation\,\cite{Law1998,Pu1999,Chang2005,Black2007,yingmeiliu2009,Hamley2012,hoang2016parametric}
for the spin component density profiles, 
atomic spin-mixing dynamics in a magnetic field as depicted in Fig.\,\ref{fig:scheme} is governed 
by the Hamiltonian
$ \hat H = \hat{H}_{0} + \hat{H}_{\rm B} + \omega_c\hat{c}^\dagger\hat{c}+
 \hat{H}_{\rm AL}$ ($\hbar=1$)\,\cite{Law1998,Pu1999},
with $\hat{H}_{0}=({c}/{2N})[(\hat N_1-\hat N_{-1})^2+(2\hat N_0-1)(\hat N_1+\hat N_{-1})
+2(\hat a_1^\dag\hat a_{-1}^\dag\hat a_0\hat a_0+{\rm h.c.})]$ 
resulting from inherent two-body $s$-wave spin-exchange at rate $c$
and $N$ the total number of atoms. 
$\hat a_{m = 0,\pm 1}$ denotes annihilation operator for condensed atoms in 
spin component $|F=1, m\rangle$ and $\hat N_{m}=\hat a_{m}^\dag \hat a_{m}$ 
the corresponding number operator. 
The Zeeman term inside a magentic field is given by $\hat{H}_{\rm B}=-p\hat{F}_z+q(\hat N_1+\hat N_{-1})$ 
(wherein $\hat{F}_z\equiv\hat N_1-\hat N_{-1}$), 
with $p$ the single-atom linear Zeeman shift and $q$ the QZS
which competes with spin exchange interaction ($\propto c$) to govern system spin-mixing dynamics.
The differential laser frequency shift is set as $\omega_1-\omega_2 = 2q$, 
exactly equal to the two-atom energy deficit if spin-exchange is to occur on resonance. 
This is a necessary condition for efficient spin-mixing, especially when the bias magnetic 
field induced QZS is large and the energy matching condition is destroyed for
spin-exchange collision\,\cite{Masson2017,davis19photon, Monika21Programmable}. 
$\omega_c\hat{c}^\dagger\hat c$ is the free cavity-photon Hamiltonian 
and the atom-light interaction Hamiltonian $\hat{H}_{\rm AL}$ describes the  
multiphoton transitions shown in the lower panel of Fig.\,\ref{fig:scheme}\,(a).

In a typical ultracold \Rb{87} atom experiment, one finds $|c|\lesssim (2\pi)10\Hz$\,\cite{Chang2004,Hamley2012,Anquez2015quantum,Luo2017}.
Assuming an applied magnetic field ranging from tens to hundreds of Gausses, 
the induced Zeeman effects satisfy: $p \gg q \gg |c|$. 
Therefore, two-body collision induced spin-mixing processes in $\hat{H}_{0}$ is highly suppressed 
by the large energy mismatch between spin-exchanged states.

The atom-light interaction Hamiltonian $\hat{H}_{\rm AL}$ under rotating-wave 
approximation becomes\,\cite{carmichael1999statistical,SM}
\begin{align}
  \hat H_1&=\sum_{j=1}^N\bigg[\left(\Omega_{1} e^{i\varphi}e^{-i\Delta t}+\Omega_2 e^{-i(\Delta +2q)t} \right) |1\rangle_j\langle e_2|\nonumber\\
& \qquad + \left(\tilde{\Omega}_1 e^{i\varphi}e^{-i\Delta^\prime t}+\tilde{\Omega}_2 e^{-i(\Delta^\prime+2q) t}\right) |0\rangle_j\langle e_1| \nonumber\\
  &\qquad + g_1\hat c^\dag e^{-i(\Delta+2q-\delta)t}|0\rangle_j\langle e_2|\nonumber\\
  &\qquad+g_2\hat c^\dag e^{-i(\Delta^\prime-\delta)t}|\!-\!1\rangle_j\langle e_1|+ \text{h.c.}\bigg]\,,\label{eq:Ht_rwa}
\end{align}
where $\varphi$ is the initial phase difference between the two lasers ($\varphi=0$ hereafter).
{$\Delta\,(\Delta^{\prime})$ denotes detuning of L1 from the transition $|1\rangle\!\leftrightarrow\!|e_2\rangle\,(|0\rangle\!\leftrightarrow\!|e_1\rangle)$}
which can take values in the range of $\GHz$ and even $\THz$ between ground state manifold and akali 
atom D-line transitions in the optical range\,\cite{mottl2014roton}, and the detunings for the L2 couplings are $\Delta+2q$ and $\Delta^\prime+2q$ respectively.
With suitably locked cavity $\omega_c$, we denote $2q-\delta$ ($\delta$) as the detuning for the two-photon Raman 
transition pathways between $|0\rangle$ and $|1\rangle$ ($|\!\!-\!1\rangle$), with L1 (\,L2\,) and the cavity field 
shown respectively in the lower pannel of Fig.\,\ref{fig:scheme}\,(a).

{\it Effective Hamiltonian.---}When the detunings between optical fields  and atomic transitions are large,
{\it i.e.}, $|g_{1,2}|,\,|\Omega_{1,2}|,\,|\tilde{\Omega}_{1,2}|\ll \,\Delta^{(\prime)},\, \Delta^{(\prime)}+2q,
\,\Delta+2q-\delta,\,\Delta^\prime-\delta$,
one can neglect atomic spontaneous emission and safely eliminate the excited states $|e_1\rangle$ and $|e_2\rangle$ to 
obtain the Hamiltonian projected onto the spin-1 atomic ground state manifold\,\cite{carmichael1999statistical,Masson2017, Zhang2017c},
\begin{eqnarray}
  \hat H_2 & = & \{[\eta_1e^{i(\delta-2q)t}+\eta_2e^{i\delta t}]\hat a_0^\dag \hat a_1\hat c^\dag + \nonumber\\  
           & &   \;\;  [\tilde{\eta}_1e^{i\delta t}+\tilde{\eta}_2e^{i(2q+\delta) t}]\hat a_{-1}^\dag \hat a_0\hat c^\dag + \text{h.c.} \}\label{eq:H2_es_elimination}\,,
\end{eqnarray}
where the two-photon Raman coupling strengths satisfy $\eta_1\approx\tilde{\eta}_1,\,\eta_2\approx\tilde{\eta}_2$,
after ac Stark shifts induced by light fields are neglected for the three ground-state levels\,\cite{SM}. 

Since the parameters $2q\pm\delta$ and $\delta$ are larger than $N|\eta_{1,2}|$ in typical experiments, 
the cavity-assisted Raman coupling between different ground states are far off-resonant,
except for the four-photon resonance pathways (presented in Fig.\,\ref{fig:scheme}\,(b)) 
accompanied by two atom {spin-exchange} that conserves the total z-component angular momentum
$|0\rangle+|0\rangle \rightleftharpoons |1\rangle +|\!-\!1\rangle$\,\cite{Chang2004}.
{We take $\delta=3q/2$,} the Hamiltonian $\hat H_2$ in Eq.\,(\ref{eq:H2_es_elimination}) 
then reduces to be time-periodic with fundamental frequency $2q-\delta=q/2$. 
Since $q/2$ is large compared to the magnitudes of matrix 
elements of the Hamiltonian, a time-independent effective Hamiltonian 
can be derived by adopting the high-frequency expansion\,\cite{SM,Eckardt2016}.
The Raman transition pathways\,(L1/L2 plus cavity mode in Fig.\,\ref{fig:scheme})
with large two-photon detunings would 
only virtually excite the cavity mode if one starts from a cavity in vacuum state\,\cite{Zheng2000,Zhang2017c}. 
The condition of $\delta=q$ is avoided in order to circumvent simultaneous 
cavity-photon-pair-creation ($\hat{c}^\dagger\hat{c}^\dagger$-term) in the four-photon resonance, 
although such processes can be used to generate multiphoton pulses\,\cite{parkins21proposal}.
Therefore we substitute the cavity mode operators by 
$\langle \hat c\hat c^\dag\rangle=1$ approximately, and neglect other cavity operators 
that shall remain negligibly small.
Finally, we obtain the time-independent effective Hamiltonian\,\cite{SM},
 \begin{equation}
   \hat H_{\rm eff} = ({\tilde{c}}/{N})(\hat a_1^\dag \hat a_{-1}^\dag \hat a_0 \hat a_0+{\hat a_0^\dag \hat a_{0}^\dag \hat a_1 \hat a_{-1}} )- \tilde{q}_0\hat N_{-1} \hat N_0,
   \label{eq:Heff}
 \end{equation}
with effective spin-mixing rate coefficient $\tilde{c}=-{2\sqrt{3}N\eta^2}/{(3q)}$ 
and $\tilde{q}_0\sim\mathcal{O}({\tilde{c}}/{N})$ when $\eta^2 = \eta_1^2=\eta^2_2/3$ are taken.
We have neglected a minute quadratic Zeeman term $-\tilde{q}_0\hat N_0$ in deriving Eq.\,(\ref{eq:Heff})
as detailed in the supplemental material\,\cite{SM}, 
since $|\tilde{q}_0|\ll|\tilde{c}|$ as a result of $N\gg 1$ implies
spin-mixing dynamics is hardly modified.
The spin-mixing term $(\hat a_1^\dag \hat a_{-1}^\dag \hat a_0 \hat a_0+{\rm h.c.})$ 
in the effective Hamiltonian Eq.\,(\ref{eq:Heff}) thus is engineered based on the intuitive
two off-resonant Raman pathways, as depicted in Fig.\,\ref{fig:scheme}(b).
The remaining density interaction is proportional to $\hat N_{-1}\hat N_0$, which can be 
regarded as a $\hat N_{-1}$-dependent QZS and may be neglected if the intermediate states have 
vanishing population in $|\!-\!1\rangle$ during spin-mixing. 

\begin{figure}
  \centering
  \includegraphics[width=1.0\columnwidth]{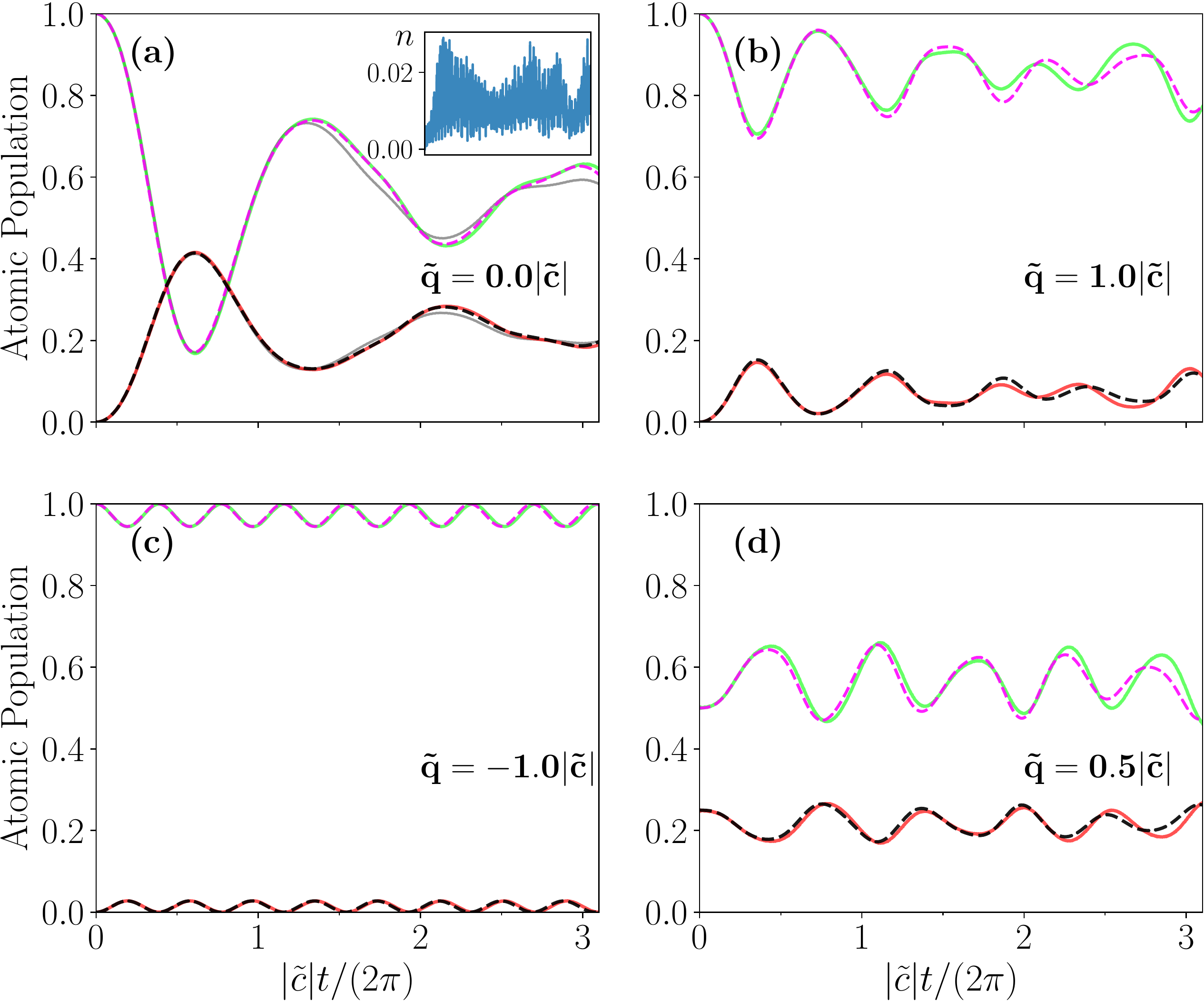}
  \caption{Spin state atomic populations during quench-$\tilde{q}$ dynamics. 
  Solid (dashed) lines denote simulations with $\hat{H}_2$ ($\hat{H}_{\rm eff}$), 
  shown in cyan-solid (magenta-dashed) and red-solid (black-dashed) lines for $n_0(t)$ and $n_1(t)$, respectively.
  (a-c) for evolution from initial state $|\Psi_0\rangle=|0,N,0\rangle\!\otimes\!|0\rangle_c$ with
  effective QZS $\tilde{q}=0|\tilde{c}|, 1.0|\tilde{c}|$, and $-1.0|\tilde{c}|$, respectively. 
  Grey-solid lines in (a) are results with cavity dissipation $\kappa=2|\tilde{c}|$ included, 
  and the inset of (a) shows photon population $n(t)$ of the cavity mode during quench-$\tilde{q}$ dynamics.
  (d) for $|\Psi_0\rangle = |N/4,N/2,N/4\rangle\!\otimes\!|0\rangle_c$ and $\tilde{q}=0.5|\tilde{c}|$. 
Other parameters used for numerical simulation are  $N=20$ and $q=6N\eta$.}
\label{fig:quench_dynamics}
\end{figure}

{\it Tunability.---}The effective cavity-mediated spin-mixing rate coefficient 
per atom  $|\tilde{c}|/N\propto{\eta^2}/{q}$ is directly determined by the intensities and detunings of the pump laser fields, 
while the single-atom QZS ($q$) remain tunable as in previous implementation by changing the 
applied magnetic or near-resonant microwave dressing fields\,\cite{YLiu14pra,Hamley2012}.
These two key parameters governing atomic spin-mixing dynamics are therefore tunable experimentally.
The sign of $\tilde{c}$ could also change to be positive, {\it i.e.}, becoming antiferromagnetic-like 
when the two Raman couplings assume an alternative configuration of detunings.  
This could be used to simulate dynamics of anti-ferromagnetically interacting spin-1 BEC\,\cite{Sun2017singlet,Masson2019singlet,Qu20PRL}, 
in a ferromagnetic one like \Rb{87} atoms.

\par  
In pioneering experimental studies\,\cite{YLiu14pra, Qu20PRL, hoang2016parametric,Luo2017}, 
microwave dressing fields are implemented to augment the effective tuning of QZS from positive to negative, 
however with tunable range limited by the available power of microwave field.
In the scheme we present, effective control of QZS can be directly accomplished, 
without requiring microwave or optical dressing,
but by a slight detuning from the four-photon resonance 
in Fig.\,\ref{fig:scheme}(b),  namely by taking the differential 
laser frequency $\omega_1-\omega_2 = 2(q-\tilde{q})$.
An effective quadratic Zeeman term $\mathcal{H}_{\rm QZS} = -\tilde{q}\hat{N}_0$ in addition to the effective 
Hamiltonian $\hat{H}_{\rm eff}$ in Eq.(\ref{eq:Heff}) would emerge.
The deviation of $2\tilde{q}$ is so small ($|\tilde{q}|\ll q$) that it hardly 
modifies the effective spin-mixing rate coefficient $\tilde{c}$, but the magnitude of $\tilde{q}$ can be 
easily controlled to be on the same order of $|\tilde{c}|$, {\it i.e.} $\tilde{q}\sim |\tilde{c}|$. 
This tunable effective QZS constitutes a key contribution of this work which complements the synthesized
spin-exchange interaction already discussed\,\cite{Masson2017,davis19photon,Monika21Programmable}.
It will enable the realizations of different quantum phases as well as flexible fast dynamics control in spinor 
atomic BEC for a variety of research topics\,\cite{Lucke2011,Anquez2015quantum,Luo2017,guo21faster}.

Futhermore, the two pump laser beams L1 and L2 in Fig.\,\ref{fig:scheme} can be derived
from a single laser by an acousto-optic modulator (AOM). 
Experimentally the difference between $\omega_1/2\pi$ and $\omega_2/2\pi$ can be well 
controlled to a high precision at the order of one Hertz. 
Therefore the frequency difference $\omega_1-\omega_2 = 2q$ ($\sim$\,\MHz) between L1 and L2 and the 
effective QZS $\tilde{q}$ ($\sim$\,\kHz) can both be precisely tuned.

For estimation of parameters and numerical simulations, we use \Rb{87} atoms with 
$|c|\lesssim (2\pi)\,10\,\Hz$ for $N\in[10^3,\,10^5]$ as in 
current BEC experiments\,\cite{hoang2016parametric,peise2015satisfying,Klempt2016PRL,Oberthaler2016PRL,Luo2017,Zou201715105}, 
the linear and quadratic Zeeman shifts at bias magnetic field $B_z$ are given by
$(p,q)=2\pi\,(0.70$\,$B_z\,\MHz/{\rm G}$,\;$71.6\,B_z^2\,\Hz/{\rm G^2})$\,\cite{Luo2017}. 
At a high $B_z= 80\,$G, $(p,q)\approx 2\pi(5.6, 0.46)\,\MHz$, 
thus one can safely neglect the inherent spin-exchange interaction ($\propto c$)
as $q\gg |c|$, and spin-mixing becomes highly suppressed by the energy mismatch $q$ per atom. 
For a cloud of $N=20$ atoms inside an optical cavity, we can take
$g = (2\pi)\,1.0\,{\rm MHz}$\,\cite{Esslinger2007,Baumann2010,Mottl2012,Landig2016,Leonard2016, Monika21Programmable}, 
assume $g_1=g_2\equiv g$ and the Rabi frequencies for the two pump lasers are 
$\Omega_1=\Omega_2\equiv\Omega=(2\pi)\,40\,\MHz$, 
the detunings for the two lasers from \Rb{87} atom D-line transition are
taken respectively as $\Delta\approx\Delta^\prime \approx (2\pi)\,21\,\GHz$\,\cite{Monika21Programmable}.
The two-photon Raman coupling strength then reduces to $\eta \approx {2g \Omega}/{\Delta} \approx (2\pi)\,3.8\, {\rm kHz}$, 
and the effective spin-mixing rate becomes $|\tilde{c}|\approx (2\pi)\,730\, {\rm Hz}$, 
which is many orders of magnitude larger than $|c|$ from inherent spin-exchange collisions.

\begin{figure}
  \centering
  \includegraphics[width=1.\columnwidth]{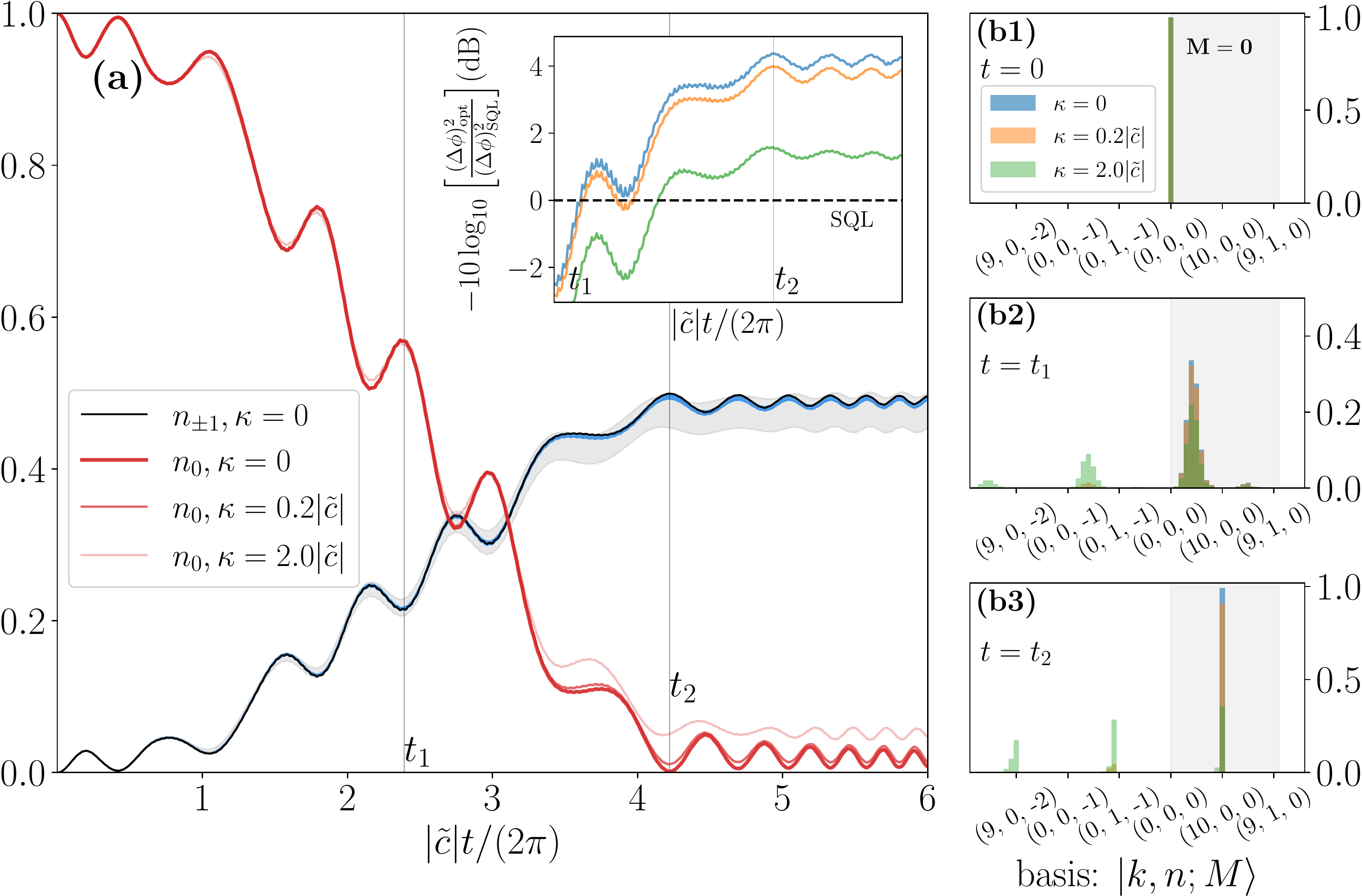}
  \caption{(a) Adiabatic preparation of atomic twin-Fock state from a linear-$\tilde{q}$ driving
  starting with an initial polar state $|N_1, N_0, N_{-1}\rangle=|0,N,0\rangle$, of zero magnetization $M=0$.
  Red- and black-solid lines denote $n_0$ and $n_{\pm1}$ defined in the main text (for $\kappa=0$), respectively.
  The shaded area, with upper [$n_{-\!1}(t)$] and lower [$n_1(t)$] borderlines
  surrounding $n_{\pm1}$ at $\kappa=0$ (black solid line),  
  measures the deviation from $M=0$ [$n_1=n_{-\!1}$] due to photon loss
  at rate $\kappa=0.2|\tilde{c}|$ (blue shaded) or $2.0|\tilde{c}|$ (gray shaded), respectively.
  Inset: metrology gain of optimal phase sensitivity [$(\Delta\phi)_{\rm opt}$] 
  beyond SQL [$(\Delta\phi)_{\rm SQL}$] (dashed line).
  (b1-b3) The probability distributions of cavity-atom state $\rho(t)$ in the Fock 
  basis $|k,n; M\rangle$ at different time $t=0$, $t_1$, and $t_2$ [labeled by gray vertical lines in (a)] respectively.
  $N=20$, $q=6N\eta$, and the Hilbert space is truncated at $|M|\leq2$ with $\max{n}=1$.
   }\label{fig:sweep_kappa}
\end{figure}

\section{Numerical simulations}
We now confirm the validity of $\hat{H}_{\rm eff}$ 
in Eq.\,(\ref{eq:Heff}) with a tunable effective QZS ({\it i.e.} $\tilde{q}$) by numerically simulating the quench-$\tilde{q}$ dynamics
following earlier experimental protocols\,\cite{Lucke2011,Klempt2016PRL}.
In the Fock state representation, with atom state $|\psi\rangle=|N_1, N_0, N_{-1}\rangle$ and the cavity state $|n\rangle_c$, 
the complete  basis state for the system becomes $|\Psi\rangle = |\psi\rangle\otimes|n\rangle_\text{c}$
specified by $N_1,N_0, N_{-1}$, and $n$. 
Assuming atoms initially reside in the polar state $|\psi_0\rangle=|0,N,0\rangle$ 
with zero magnetization, which is easy to prepare experimentally\,\cite{Luo2017,hoang2016parametric}, 
and the cavity is empty in vacuum state $|0\rangle_{c}$,
the atomic population $n_{m}(t)=\langle \hat N_{m}\rangle/N$ ($m=0, \pm 1$) is 
simulated by evolving the initial state $|\Psi_0\rangle=| \psi_0\rangle \otimes|0\rangle_c$ 
under both $\hat H_2(t)$ in Eq.\,(\ref{eq:H2_es_elimination}) and the effective Hamiltonian $\hat{H}_{\rm eff}$ in Eq.\,(\ref{eq:Heff}). 
The two results of quench-$\tilde{q}$ dynamics are compared in Fig.\,\ref{fig:quench_dynamics}.
In Fig.\,\ref{fig:quench_dynamics}\,(a), 
with initial atomic polar state, 
the effective Hamiltonian $\hat H_\text{eff}$ is found to work quite well over an extended 
time scale with respect to the characteristic spin-mixing time scale $1/|\tilde{c}|$. 
The inset of Fig.\,\ref{fig:quench_dynamics}\,(a) shows that population of the cavity mode 
remains negligibly small ($n\!\ll\!1$) during the time evolution, supporting our assumption 
that cavity mode is only virtually excited and thus our scheme is found to be immune to photon loss. 
By adjusting the laser frequency difference between L1 and L2, we effectively 
tune the QZS by $\tilde{q}$. Figures\,\ref{fig:quench_dynamics}\,(b) and (c) show evolutions from the same state $|0,N,0\rangle\otimes|0\rangle_c$ 
but at different effective QZS $\tilde{q}=\pm |\tilde{c}|$. 
Comparisions are also performed for a different initial state 
$|\Psi_0\rangle=| N/4, N/2,N/4\rangle\otimes|0\rangle_c$ in Fig.\,\ref{fig:quench_dynamics}\,(d) 
whose results again support the validity of the effective Hamiltonian in Eq.\,(\ref{eq:Heff}).

The effects of photon loss from cavity can be included by
employing master equation for the complete cavity-atom state $\rho$,
\begin{eqnarray}
{\partial_t \rho(t)} = -i[\hat {H}_2(t),\rho] +({\kappa}/{2})\mathcal{D}(\hat{c}, \rho)\;,\label{eq:mastereq}
\end{eqnarray}
where the Lindblad term 
$\mathcal{D}(\hat{c}, \rho) = 2\hat{c}\rho\hat c^\dagger-\hat{c}^\dagger\hat{c}\rho-\rho\hat{c}^\dagger\hat{c}$ 
describes dissipative processes associated with cavity loss at rate $\kappa$.
We can use an ultranarrow-band optical cavity with $\kappa/(2\pi)$ at the order of $\sim {\rm kHz}$\,\cite{Kebler2014, Klinder15pnas,georgakopoulos2018modeling},
hence we choose $\kappa/|\tilde{c}|=0, 0.2, 2.0$ for the following simulations at $N=20$.
We find that evolutions of atomic populations are hardly modified when cavity dissipations are included 
as shown in Fig.\,\ref{fig:quench_dynamics}\,(a) by grey solid lines ($\kappa=2|\tilde{c}|$).

To emphasize the utility of controlling both $\tilde{c}$ and $\tilde{q}$ in the present scheme, 
we simulate a dynamic driving-$\tilde{q}$ protocol as shown in the inset of Fig.\,\ref{fig:sweep_kappa}, 
which is implemented here for adiabatic preparation of metrologically useful quantum entangled states\,\cite{Luo2017, Zou201715105}.
The effective QZS $\tilde{q}$ is swept linearly by scanninng frequency difference of the two pump lasers, 
from polar to twin-Fock phases of the instaneous effective Hamiltonian. 
Figure \ref{fig:sweep_kappa}\,(a) shows the atomic population transfer from $|0\rangle$
to the spin $|\!\pm\!1\rangle$ states largely follows the driving-$\tilde{q}$, 
and almost perfectly prepares the desired twin-Fock state at the moment of $t\!=\!t_2$ as clearly revealed by the state distribution 
in the Fock basis $|k,n; M\rangle$ in Fig.\,\ref{fig:sweep_kappa}\,(b3).
A short hand for the basis state is now indexed according to $|k,n; M\rangle\equiv|N_1, N_0, N_{-1}\rangle\otimes|n\rangle_c$, 
with $N_1=k$, $N_0=N+M-2k-n$, and $N_{-1}=k-M+n$, where $N=N_1+N_0+N_{-1}$ and $M=N_1-N_{-1}+n$
are good quantum numbers in the absence of dissipation for initial $M=0$.
A near unit peak centered at the target twin-Fock state $|10,0;0\rangle$ results even in the presence of dissipation. 
The inset of Fig.\,\ref{fig:sweep_kappa}\,(a) shows phase sensitivity 
of the prepared twin-Fock state when fed into a Ramsey interferometer\,\cite{SM,Luo2017}.
We find an entanglement-enhanced phase sensitivity [$(\Delta\phi)_{\rm opt}$] at $t_2$ of about 
$4.3$\,dB beyond the standard quantum limit (SQL) [$(\Delta\phi)_{\rm SQL}=1/\sqrt{N}$] at $\kappa=0$,
and the enhancement reduces to $1.6$\,dB at $\kappa=2|\tilde{c}|$, 
implicating a favorable robustness of the present scheme for operational metrology gain.
Photon loss tends to polarize atoms into $M\leq 0$, exemplified by the shifting distributions 
in Fig.\,\ref{fig:sweep_kappa}\,(b) out of the initial $M=0$ subspace 
from $t=0\rightarrow t_1\rightarrow t_2$.
Such atomic polarization mainly arise from the intrinsic asymmetry of our scheme by 
the presence of a significant QZS, which is not considered in early 
studies\,\cite{Masson2017,davis19photon,Monika21Programmable}.

\section{Conclusions}
In conclusion, we have proposed an efficient scheme for controlled spin-mixing dynamics
based on tuning of both the spin-mixing rate and the competing QZS by changing pump lasers parameters. 
The tuned interaction occurs on a much faster time scale than inherent spin-exchange dynamics,
and the synthesized spin-spin interaction as well as the effective QZS are essentially independent 
of the inherent atomic collision properties, therefore can be generalized to other 
atomic species, such as atoms with higher spins, {alkali metals, or atomic mixtures}.
We hope this work will open the doors to more tunability in cold atom spin-spin interactions 
and their dynamics controls to enrich future experimental studies.

\section*{Acknowledgements}
We ackowledge helpful discussions with Qi Liu and Xinwei Li.
Numerical simulations are performed with QuTiP\,\cite{johansson2012qutip,johansson2013qutip}.


\begin{appendix}
\section{Adiabatic elimination of excited states}\label{sec:appendix_elimination}
Inside a large bias magnetic field, the inherent contact interactions between atoms 
  are highly suppressed by the quadratic Zeeman shift (QZS). 
  The atom-cavity Hamiltonian becomes
  $\hat H\approx \hat H_B + \omega_c\hat{c}^\dagger\hat{c}+\hat{H}_{\rm AL}\,,
  $
  with $\hat{H}_B$ and $\hat{H}_{\rm AL}$ given in the main text.
  In the rotating frame defined by $\hat U=\exp\{i(\hat{H}_B+\omega_c \hat c^\dag \hat c)t\}$, 
  the Hamiltonian in the new frame becomes
  $\tilde{\hat{H}} = \hat U \hat H\hat U^\dag+(i\partial_t\hat U)\hat U^\dag$, 
  which is denoted by $\hat{H}_1$ in the main text after rotating-wave approximation.
  
  The large single- and two-photon detunings considered in this work ensures
  negligible occupation on atomic excited states, 
  therefore adiabatic elimination of excited states is appropriate\,\cite{carmichael1999statistical,sorensen12effective}.
  
  We find after tedious calculations
  \begin{eqnarray}
    \hat H_2 &=&\{[\eta_1e^{-i(2q-\delta)t}+\eta_2e^{i\delta t}]\hat a_0^\dag \hat a_1\hat c^\dag\\
    &&+[\tilde{\eta}_1e^{i\delta t}+\tilde{\eta}_2e^{i(2q+\delta) t}]
    \hat a_{-1}^\dag \hat a_0\hat c^\dag +\text{H.c.}\}+\hat H_\text{Stark}\nonumber\,,
  \end{eqnarray}
  where $\eta_{1,2}$ ($\tilde{\eta}_{1,2}$) denote the cavity-assisted two-photon Raman coupling strengths defined by
  \begin{eqnarray*}
       \eta_1&=&{g_1\Omega_1}\left(\frac{1}{\Delta}+\frac{1}{\Delta+2q-\delta}\right),\\
        \eta_2&=&g_1\Omega_2\left(\frac{1}{\Delta+2q}+\frac{1}{\Delta+2q-\delta}\right)\,,\\
       \tilde{\eta}_1&=&{g_2\tilde{\Omega}_1}\left(\frac{1}{\Delta^\prime}+\frac{1}{\Delta^\prime-\delta}\right),\\
        \tilde{\eta}_2&=&g_2\tilde{\Omega}_2\left(\frac{1}{\Delta^\prime+2q}+\frac{1}{\Delta-\delta}\right)\,.
    \end{eqnarray*}
  The Stark shift $\hat H_\text{Stark} = [(\frac{\Omega_1^2}{\Delta}+\frac{\Omega_2^2}{\Delta+2q})\hat a_1^\dag a_1
  +(\frac{g_1^2\hat c^\dag\hat c}{\Delta+2q-\delta}+\frac{\tilde{\Omega}_1^2}{\Delta^\prime}
  +\frac{\tilde{\Omega}_2^2}{\Delta^\prime+2q})\hat a_0^\dag \hat a_0+
  (\frac{g_2^2\hat c^\dag\hat c}{\Delta^\prime-\delta})\hat a_{-1}\hat a_{-1}]
  + [\Omega_1\Omega_2e^{i2qt}(\frac{1}{\Delta}+\frac{1}{\Delta+2q})\hat a_1^\dag \hat a_1 
  + \tilde{\Omega}_1\tilde{\Omega}_2e^{i2qt}(\frac{1}{\Delta^\prime}
  +\frac{1}{\Delta^\prime+2q})\hat a_0^\dag \hat a_0+ \text{h.c.}]$
  induced by light fields is neglected in Eq.(3) of the main text for three ground-state levels.
  This term can be absorbed into the initial linear and quadratic Zeeman shifts in $\hat{H}_B$.
  In fact, it is much smaller than the Zeeman shifts for resonable sized bias magnetic field. 
  
  \section{Time-independent Hamiltonian with Floquet-Magnus expansion}\label{append:floquet}
  Here we use a simple approach to derive the effective Hamiltonian $\hat{H}_{\rm eff}$ given in the main text. 
  
  For the time periodic Hamiltonian,
  \begin{eqnarray}
   \hat{H}_2(t)
    &=&[(\eta_1 e^{-i\omega t}+\eta_2 e^{i3\omega t})\hat a_0^\dag\hat a_1\hat c^\dag\nonumber\\
     && +(\eta_1 e^{i3\omega t}+\eta_2 e^{i7\omega t})\hat a_{-1}^\dag\hat a_0\hat c^\dag+\text{h.c.}]\,,
   \end{eqnarray}
  with $\omega=q/2$, we can carry out the Floquet-Magnus expansion\,\cite{Eckardt2016} and keep terms till the order of ${1}/{\omega}$.
  This yields the time-independent Hamiltonian,
   \begin{eqnarray}
      \hat H_{\rm eff}&=&\frac{1}{\omega}[\hat V_1, \hat V_{-1}]+\frac{1}{3\omega}[\hat V_3, \hat V_{-3}]+\frac{1}{7\omega}[\hat V_7, \hat V_{-7}]\,,
   \end{eqnarray}
   where $[\hat A, \,\hat B]=\hat A\hat B-\hat B\hat A$ being the commutator of operators $\hat A$ and $ \hat B$,
   and we have 
  \begin{eqnarray*}
    \hat V_1 &=& \eta_1 \hat a_1^\dag \hat a_0 \hat c, \quad
    \hat V_3 = \eta_2 \hat a_0^\dag \hat a_1\hat c^\dag+\eta_1 \hat a_{-1}^\dag \hat a_0\hat c^\dag ,\\
    \hat V_{7}&=&\eta_1 \hat{a}_{-\!1\!}^\dag \hat{a}_0 \hat{c}^\dag, \quad
    \hat V_{m}=0\quad \text{for}\, m = \text{other integers}.
  \end{eqnarray*}
  
   It is straightforward to work out all the terms, and we find
   \begin{eqnarray*}
  {[\hat{V}_1, \hat{V}_{\!-\!1}]}&=&\eta_1^2 (\hat{a}^\dag_1 \hat{a}_1 \hat{c}\hat{c}^\dag -\hat{a}_0^\dag \hat{a}_0 \hat{c}^\dag\hat{c}) 
    +\eta_1^2 \hat{a}_1^\dag \hat{a}_0^\dag \hat{a}_0\hat{a}_1[\hat{c}, \hat{c}^\dag],\\        
  {[\hat{V}_3, \hat{V}_{\!-\!3}]}&=&(\eta_1^2 \hat a_0^\dag \hat a_{-1}^\dag \hat a_{-1} \hat a_0
  +\eta_2^2\,\hat a_0^\dag \hat a^\dag_1\hat a_1\hat a_0)\cdot[\hat{c}^\dag, \hat{c}]\\
  &&+\eta_1\eta_2(\hat a_0^\dag \hat a_0^\dag \hat a_1\hat a_{-1}+\hat a_{-1}^\dag\hat a_1^\dag \hat a_0\hat a_0)\cdot[\hat c^\dag, \hat{c}]\\
  &&+\,\eta_1^2(\hat a_{-1}^\dag\hat a_{-1}\hat{c}^\dag\hat{c}-\hat a_0^\dag\hat a_0\hat{c}\hat{c}^\dag)\\
  &&+\eta^2_2(\hat a_0^\dag \hat a_0\hat{c}^\dag \hat{c}-\hat a_1^\dag \hat a_1\hat{c} \hat{c}^\dag),\\  
  {[V_7, V_{\!-\!7}]}&=&\eta_2^2\,\hat a_{-1}^\dag \hat a_0^\dag \hat a_0 \hat a_{-1}\cdot[\hat{c}^\dag, \hat{c}]\\
      &&+\eta^2_2\,(\hat{a}_{-1}^\dag \hat a_{-1} \hat{c}^\dag \hat{c} -\hat a_0^\dag\hat a_0 \hat{c} \hat{c}^\dag)\,,
  \end{eqnarray*}
  which give
   \begin{eqnarray*}
  \hat H_{\rm eff}\cdot \omega &=&
    (\frac{1}{3}\eta_2^2-\eta_1^2)\hat a_1^\dag \hat a_0^\dag \hat a_0\hat a_1 \cdot[\hat c^\dag,\,\hat c]\\
    &&+(\frac{1}{3}\eta_1^2+\frac{1}{7}\eta_2^2)\hat a_{-1}^\dag\hat a_0^\dag \hat a_0 \hat a_{-1}\cdot[\hat c^\dag,\,\hat c]\\
      &&+\frac{1}{3}\eta_1\eta_2 (\hat a_0^\dag\hat  a_0^\dag \hat  a_{1}\hat a_{-1}+\hat a^\dag_1 \hat a^\dag_{-1}\hat a_0 \hat a_0)\cdot[\hat c^\dag, \hat c]\\
     &&+ (\eta_1^2-\frac{1}{3}\eta_2^2)(\hat a_1^\dag \hat a_1 \hat c \hat c^\dag-\hat a_0^\dag \hat a_0\hat c^\dag \hat c) \\
     && +(\frac{1}{3}\eta_1^2+\frac{1}{7}\eta_2^2)
     (\hat a^\dag_{-1}\hat a_{-1}\hat c^\dag\hat c-\hat a_0^\dag \hat a_0 \hat c \hat c^\dag)\,.
   \end{eqnarray*} 
Substituting into $\hat{c}^\dag \hat{c}=0$, $\hat{c}\hat{c}^\dag=1$, and taking $\eta_2^2=3\eta_1^2=3\eta^2$, 
the above reduces to the effective Hamiltonian,
  \begin{equation}
    \hat H_\text{eff}=\frac{\tilde{c}}{N}(\hat a_0^\dag\hat a_0^\dag \hat a_{1}\hat a_{-1} + {\rm h.c.})-\tilde{q}_0 \,\hat a_{-1}^\dag\hat a_0^\dag \hat a_0 \hat a_{-1}-\tilde{q}_0\,\hat a_0^\dag \hat a_0\,,
  \end{equation}
  where ${\tilde{c}}=-\sqrt{3}{N\eta^2}/{(3\omega)}$ 
  and $\tilde{q}_{0}={16\eta^2}/({21\omega})={16\sqrt{3}}|\tilde{c}|/({21N})$.
  The residual density-density interaction term $\propto \hat a^\dag_{-1}\hat a_0^\dag\hat a_0\hat a_{-1}$ 
  does not appreciably modify the spin-mixing dynamics.

  Note that $\tilde{c}$ may be positive if we choose an alternative cavity frequency condition to 
  maintain an opposite sign of two-photon detuning, 
  thereby rendering anti-ferromagnetic atomic spin-exchange interaction as in $^{23}$Na atoms.

\section{Tunability of the effective quadratic Zeeman shift}\label{sec:app:tune_q}
We consider $(\omega_1,\omega_2)\rightarrow(\omega^\prime_1,\omega^\prime_2)=(\omega_1+\tilde{q}/2,\omega_2+5\tilde{q}/2)$, 
which gives $\omega_1^\prime-\omega^\prime_2=2(q-\tilde{q})$, with $|\tilde{q}|$ ($\ll q$) a small deviation from $2q$. 
The time-periodic Hamiltonian then takes the form
\begin{eqnarray*}
  \hat H(t) &=& \{[\eta_1 e^{-i({q}/{2}+{\tilde{q}}/{2}) t}
  +\eta_2 e^{i({3q}/{2}-{5\tilde{q}}/{2}) t}]
  \hat a_0^\dag\hat a_1\hat c^\dag \\
  &&+[\eta_1 e^{i({3q}/{2}-{\tilde{q}}/{2}) t}
  +\eta_2 e^{i({7q}/{2}-{5\tilde{q}}/{2}) t}]a_{-1}^\dag a_0 c^\dag+ \text{h.c.} \}\,,
\end{eqnarray*}
we now change to work in the rotating frame defined by $\hat U^\prime= e^{i\tilde{q}\hat a_0^\dag\hat a_0 t}$ and find
\begin{eqnarray*}
  \tilde H &=& \{[\eta_1 e^{-i(\frac{q}{2}-\frac{\tilde{q}}{2}) t}
  +\eta_2 e^{i(\frac{3q}{2}-\frac{3\tilde{q}}{2}) t}] \hat a_0^\dag\hat a_1\hat c^\dag \\
  &&+[\eta_1 e^{i(\frac{3q}{2}-\frac{3\tilde{q}}{2}) t}
  +\eta_2 e^{i(\frac{7q}{2}-\frac{7\tilde{q}}{2}) t}]a_{-1}^\dag a_0 c^\dag+ \text{h.c.} \}-\tilde{q}\hat a_0^\dag\hat a_0\,.
\end{eqnarray*}
Following the same Floquet-Magnus approximation, we arrive at 
\begin{eqnarray}
  \hat H_\text{eff} &=& \frac{\tilde c}{N}(\hat a_0^\dag \hat a_0^\dag \hat a_{-1}\hat a_{-1}+ {\rm h.c.})
  -\tilde q_0\,\hat a^\dag_{-1}\hat a_0^\dag\hat a_0\hat a_{-1}\nonumber\\
  &&-\tilde q_0\hat a^\dag_0\hat a_0-\tilde{q}\hat a^\dag_0\hat a_0,
\end{eqnarray}
where $\tilde{c}=-{\sqrt{3}N\eta^2}/{(3\omega)}$, $\tilde{q}_0={16\eta^2}/{(21\omega)}$, and $\omega={(q-\tilde{q})}/{2}$.

Therefore $\tilde{q}$ indeed behaves as an effective quandratic Zeeman shift which can be easily tuned 
by changing the diffrence of two pump laser frequencies.
  
\section{Phase sensitivity}
  The optimal phase sensitivity\,\cite{Luo2017} 
  \begin{equation}
    (\Delta\phi)_{\rm opt}^2=
    \frac{V_{xz}+2\Delta\hat{J}_z^2\Delta\hat{J}_x^2}{4(\langle\hat{J}_x^2\rangle-\langle \hat{J}_z^2\rangle)^2},  
  \end{equation}
  where $V_{xz}=\langle(\hat{J}_x\hat{J}_z+\hat{J}_z\hat{J}_x)^2\rangle
  +\langle\hat{J}_x^2\hat{J}_z^2+\hat{J}_z^2\hat{J}_x^2\rangle
  -2\langle\hat{J}_z^2\rangle\langle\hat{J}_x^2\rangle$, 
  $\hat{J}_{i=x,y,z}$ is the collective spin operator for $N$ spin-$1/2$. 
  The expectation of obsevables are defined as $\langle\hat{O}\rangle\equiv {\rm Tr}(\hat\rho\hat O)$
  for density matrix $\hat\rho$.
  
\end{appendix}

\bibliography{mybib}

\end{document}